\documentclass[twocolumn, english,prl,preprintnumbers,amsmath,amssymb]{revtex4}
\usepackage[T1]{fontenc}
\usepackage[latin9]{inputenc}
\usepackage{amstext}
\usepackage{amsmath}
\usepackage{amsfonts}
\usepackage{amssymb}
\usepackage{mathtools}
\usepackage{graphicx}
\usepackage {bm}
\usepackage[section]{placeins}
\usepackage{color}
\makeatletter
\@ifundefined{textcolor}{}
{%
 \definecolor{BLACK}{gray}{0}
 \definecolor{WHITE}{gray}{1}
 \definecolor{RED}{rgb}{1,0,0}
 \definecolor{GREEN}{rgb}{0,1,0}
 \definecolor{BLUE}{rgb}{0,0,1}
 \definecolor{CYAN}{cmyk}{1,0,0,0}
 \definecolor{MAGENTA}{cmyk}{0,1,0,0}
 \definecolor{YELLOW}{cmyk}{0,0,1,0}
}
\def\be{\begin{equation}}
\def\ee{\end{equation}}
\def\bea{\begin{eqnarray}}
\def\eea{\end{eqnarray}}
\def\bse{\begin{subequations}}
\def\ese{\end{subequations}}

\usepackage{dcolumn}
\usepackage{bm}

\makeatother

\usepackage{babel}
\begin{document}

\preprint{\bibliographystyle{revtex4}}

\title{Quantum triple point and quantum critical endpoints in metallic magnets}

\author{D. Belitz$^{1}$ and T.R. Kirkpatrick$^{2}$}

\affiliation{
 $^{1}$Department of Physics and Theoretical Science Institute, University of Oregon, Eugene, OR 97403\\$^{2}$Institute for Physical Science and Technology, University of Maryland, College Park, MD 20742}

\date{\today}
\begin{abstract}
In low-temperature metallic magnets, ferromagnetic (FM) and antiferromagnetic (AFM) orders can exist in a single system in different parts of the 
phase diagram as a function of some control parameter. These phases can be adjacent, or exist concurrently, resulting in a phase
transition between a FM phase and an AFM one, or between a phase of concurrent FM and AFM order and either of the pure phases. 
We show that universal quantum fluctuations qualitatively alter the known phase diagrams for classical magnets: 
They shrink the region of concurrent FM and AFM order, change various transitions from second to first order, and, in the presence 
of a magnetic field, lead to either a quantum triple point where the FM, AFM and paramagnetic phases coexist, or to a quantum critical end point. 
\end{abstract}

\maketitle

Quantum phase transitions (QPTs) in metallic magnets are of great current interest. It is well established that in clean 
systems at low temperatures $T$ the paramagnetic (PM) to ferromagnetic (FM) transition is generically discontinuous or first 
order \cite{Brando_et_al_2016a}. This was predicted theoretically by Belitz, Kirkpatrick and Vojta (BKV) 
\cite{Belitz_Kirkpatrick_Vojta_1999, Kirkpatrick_Belitz_2012b}, and confirmed by numerous experiments \cite{Brando_et_al_2016a}. 
It is reconciled with the continuous or second-order nature of the classical transition
via a quantum tricritical point (QTCP; see Ref.~\onlinecite{quantum_footnote} for our use of the ``quantum'' prefix) in the $T$-dependent phase 
diagram.  In an applied magnetic field $h$ tricritical wings appear \cite{Belitz_Kirkpatrick_Rollbuehler_2005}; this also has been 
observed \cite{Brando_et_al_2016a}. At $T=0$ the wings end in a pair of critical points to be 
referred to as quantum wing critical points (QWCPs). These observations demonstrate how quantum fluctuations dramatically modify 
the classical FM transition, which is generically continuous, exists only at $h=0$, and in a magnetic field is replaced by a crossover. 

A more complex situation arises in systems where antiferromagnetic (AFM) or spin-density-wave order is observed in addition to 
FM order \cite{types_of_AFMs_footnote}. These range from relatively simple compounds, such as 
FeRh \cite{Muldawer_deBergevin_1961} and NbFe$_2$ \cite{Moroni_et_al_2009}, to more
complex Kondo-lattice systems such as CeRuPO \cite{Kotegawa_et_al_2013} and CeAgSb$_2$ \cite{Sidorov_et_al_2003}. The transition
from a pure FM to a pure AFM in clean systems is usually observed to be discontinuous, although in systems that contain substantial amount of disorder, 
such as Mn-doped Ni$_2$MnGa \cite{Enkovaara_et_al_2003}, Ba$_{0.6}$K$_{0.4}$Mn$_2$As$_2$ \cite{Pandey_et_al_2013}, and 
CaRu$_{1-x}$Mn$_x$O$_3$ \cite{Kawanaka_et_al_2009}, there may be a  
continuous transition from a pure FM phase to a phase of concurrent FM and AFM orders.

The classical theory of this transition has been discussed by Moriya and Usami (MU) \cite{Moriya_Usami_1977, Moriya_1985}. 
The relevant Landau free-energy density is
\be
f = r\, n^2 + t\, m^2 + u\, n^4 + v\, m^4 + 2w \,n^2 m^2 - h\, m\ ,
\label{eq:1}
\ee
with $n$ the AFM order parameter and $m$ the FM one. $r$, $t$, $u$, $v$ and $w$ are Landau coefficients. MU assumed that $u$, $v$, 
and $w$ were all positive; we will adopt this assumption \cite{negative_w_footnote}. Physically, $w$ is a measure of the free-energy cost 
of having simultaneous FM and AFM order. The results of MU can be summarized as follows. At $h=0$, and for $w^2 > uv$, there is a 
discontinuous transition from a pure FM phase to a pure AFM phase, see Fig.~\ref{fig:1}(a). If $w^2 < uv$, a phase of concurrent AFM and 
FM order exists (to be denoted as the FM+AFM phase); it is separated from pure 
AFM and pure FM phases by two continuous transitions, see Fig.~\ref{fig:1}(b). In the space spanned by $h$ and a control parameter that
tunes the system from FM to AFM to PM, $w^2 > uv$ leads to the AFM being confined to a dome, with part of the perimeter a
line of first-order transitions from AFM to FM, and the other part a line of second-order transitions from AFM to spin-polarized PM,
with a tricritical point (TCP) separating the two halves of the dome perimeter \cite{quantum_footnote}. 
Qualitatively, the phase diagram looks like Fig.~\ref{fig:3}(a), which includes weak quantum fluctuation effects \cite{different_AFMs_footnote}. $w^2 < uv$ leads to FM+AFM order in the
part of the dome adjacent to the FM phase, which crosses over to a mixture of AFM and spin-polarized PM in the part adjacent to the PM
phase. In this case the entire dome boundary is a line of second-order transitions.

Given the drastic modifications of the FM-to-PM transition due to quantum effects described above, it is important to study the effects
of quantum fluctuations on the more complicated free energy given by Eq.~(\ref{eq:1}), which yields a rich phase diagram already in the
classical case. This is the purpose of the present Letter.

When quantum fluctuation effects are included, a novel universal new term appears in Eq.~(\ref{eq:1}). At $T=0$ in three-dimensional
(3\,-D) and 2-D systems, respectively, it is
\bea
\delta f_{\text{D}=3} &=& {\tilde v}\,m^2 (m^2 + n^4) \ln(m^2+n^4)\ ,
\label{eq:2}\\
\delta f_{\text{D}=2} &=& -{\tilde v} m^2 (m^2+n^4)^{1/2}\ .
\label{eq:3}
\eea
In these expressions, ${\tilde v} > 0$ is a measure of the strength of the quantum fluctuation effects. The Eqs.~(\ref{eq:2}) 
and (\ref{eq:3}) can be derived by coupling the conduction electrons to the effective magnetic fields caused by the FM and 
AFM order parameters. A detailed derivation will be give elsewhere \cite{Kirkpatrick_Belitz_2017b}; here we confine ourselves 
to some plausibility arguments. In the absence of AFM order, $n=0$, the Eqs.~(\ref{eq:2}, {\ref{eq:3}) reduce to the known quantum 
effects in FMs that are included in BKV theory \cite{Brando_et_al_2016a}. The nonanalytic
term in the free energy reflects soft or massless excitations in the conduction-electron system that are rendered massive by a nonzero FM 
order parameter. The basic question is how an AFM order parameter enters this term. It cannot do so in the 
same way as the FM order parameter, since it is characterized by a large wave number. However, a pair of AFM order parameters 
can combine to couple to both the homogeneous FM order and the fermionic soft modes, which suggests that $n^2$ enters the nonanalytic 
term in the same way as $m$ does. A detailed analysis of the nature and couplings of the FM and AFM paramagnons in a metal, based on a
recent study of magnons in AFMs \cite{Kirkpatrick_Belitz_2017}, confirms this 
conjecture. Physically, the FM and AFM orders produce an effective homogeneous and staggered magnetic
field, respectively, that affects the conduction electrons. Integrating out the latter produces the nonanalytic term in the magnetic free-energy 
functional. We note that the $n^4$ term multiplying the brackets is of higher order in the order parameters and should not be taken seriously. It has no qualitative effects for the following discussion.

We now compute 3\,$D$ phase diagramsby minimizing the free-energy functional $f + \delta f$ with respect to $m$ and $n$. In 2\,-D the 
quantum effects are even stronger; this will be discussed elsewhere \cite{Kirkpatrick_Belitz_2017b}. 

\begin{figure}[t]
\includegraphics[width=8cm]{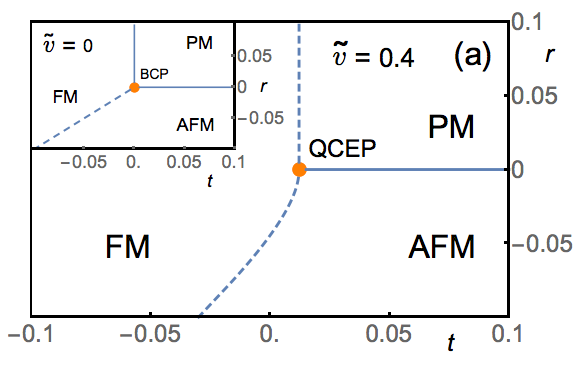}
\includegraphics[width=8cm]{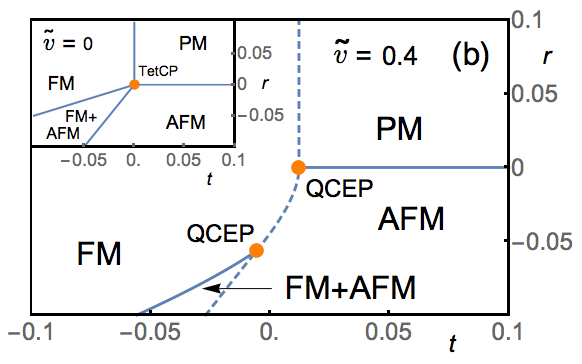}
\caption{Phase diagrams in the $r$-$t$ plane for the quantum ($\tilde v = 0.4$, main panels) and classical ($\tilde v = 0$, insets) free-energy functionals for the large-$w$ (a)
              and small-$w$ (b) case. Dashed and solid lines denote first and second-order transitions, respectively. QCEP denotes quantum 
              critical end points, BCP denotes a bicritical point, and TetCP a tetracritical point; see Refs.~\onlinecite{quantum_footnote,
              CEP_footnote} for the nomenclature used. Parameter values are $u=v=1$, $w=2$ for 
              panel (a) and $w=0.5$ for panel (b).}
\label{fig:1}
\end{figure}
{\it $r$-$t$ phase diagrams:} For the most basic phase diagram in the plane spanned by $r$ and $t$ at $h=0$, 
with all other parameters fixed, there are two possibilities:
i) A single discontinuous QPT from a pure FM state to a pure AFM state, or, ii) a
continuous QPT from a FM phase to an FM+AFM phase, followed by a discontinuous transition to a pure 
AFM phase. Which of these possibilities is realized depends on the parameter $w$ in Eq.~(\ref{eq:1}). For $w$ larger than
a critical value one has the situation shown in Fig.~\ref{fig:1}(a). There is a single transition from FM to AFM, and no FM+AFM
phase occurs. A qualitative change compared to the classical phase diagram discussed by MU, shown in the inset, is that the FM-PM
transition is first order due to the quantum fluctuations. As a result, the bicritical point (BCP) in the classical phase diagram is replaced
by a quantum critical end point (QCEP) \cite{CEP_footnote, quantum_footnote}. Quantitatively, the quantum fluctuations enlarge the FM
phase at the expense of the AFM one. For $w$ smaller than the critical value one has the situation shown in Fig.~\ref{fig:1}(b), with an 
FM+AFM phase in between the FM and AFM phases in a part of the phase diagram. However, in a qualitative change
from the classical phase diagram shown in the inset, a direct FM-to-AFM transition exists, and there are two QCEPs instead of a single 
tetracritical point (TetCP) in the classical case, and the existence of this phase is restricted to sufficiently negative values of $r$. The
latter feature can be understood from a basic feature of the free energy: Classically, for any solution with $m>0$ and $n>0$ one has
$n^2 = (-r -2w m^2)/2u$, and a relation of the same structure remains true in the quantum case. The quantum fluctuations make 
$m$ discontinuous, which implies that $n$ can be real, and the FM+AFM
solution can exist, only for sufficiently large negative $r$. In addition to the FM-to-PM transition, those from FM to AFM, and from 
FM+AFM to AFM, are all first order as a result of the quantum fluctuations;
the latter thus drastically change the nature of the phase diagram. We note that across the first-order FM-AFM 
transition in Fig.~\ref{fig:1}(b) the AFM order parameter is discontinuous, just as the FM one is. This is an example of quantum
fluctuations driving an AFM transition first order even though they couple only indirectly to the AFM order parameter.

In order to relate more directly relevant to experiments, consider a control parameter $p$
on which both $t$ and $r$ depend. Changing $p$ will thus map out a path in the $t$-$r$ plane. In an actual experiment, $p$ is often,
but not necessarily, realized by hydrostatic pressure \cite{Brando_et_al_2016a}. For simplicity we will consider only
linear paths:
\be
r(p) = r_0 + (r_1 - r_0)p \quad,\quad t(p) = t_0 + (t_1 - t_0)p \ .
\label{eq:4}
\ee

{\it $w$-$p$ phase diagram:}
To illustrate the qualitative dependence on $w$ we show the phase diagram in the $w$-$p$ plane in Fig.~\ref{fig:2} for paths that
start in the FM phase and end in the AFM phase. Comparing with
the classical case we see again that the quantum effects shrink the FM+AFM phase, change the transition
from the latter to the AFM from second to first order, and change the classical BCP to a QCEP. In addition, they lead to a 
pronounced asymmetry of the phase diagram, whereas the classical one is symmetric with respect to $p=0.5$ for the path chosen.
\begin{figure}[t]
\includegraphics[width=8cm]{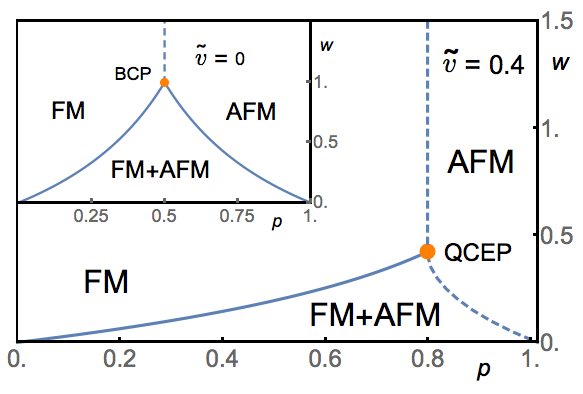}
\caption{Phase diagrams in the $w$-$p$ plane for the quantum (main panel) and classical (inset) free-energy functionals with $u=v=1$. 
              The quantum effects lead to a strong asymmetry of the phase diagram, replace one of the second-order transitions with a first-order
              one, and the classical bicritical point (BCP) with a quantum critical end point (QCEP, see Refs.~\onlinecite{quantum_footnote}, 
              \onlinecite{CEP_footnote}). 
              $p$ parameterizes a linear path in the $r$-$t$ plane, Eq.~(\ref{eq:4}) and Fig.~\ref{fig:1}, with $r_0=0$,
              $r_1=-0.05$, $t_0=t_1-0.05$. For the main panel, $t_1$ is the $t$-value corresponding to the QCEP located at $r=0$ in Fig.~\ref{fig:1};
              for the inset, $t_1=0$.}
\label{fig:2}
\end{figure}
\begin{figure}[t]
\includegraphics[width=7.4cm]{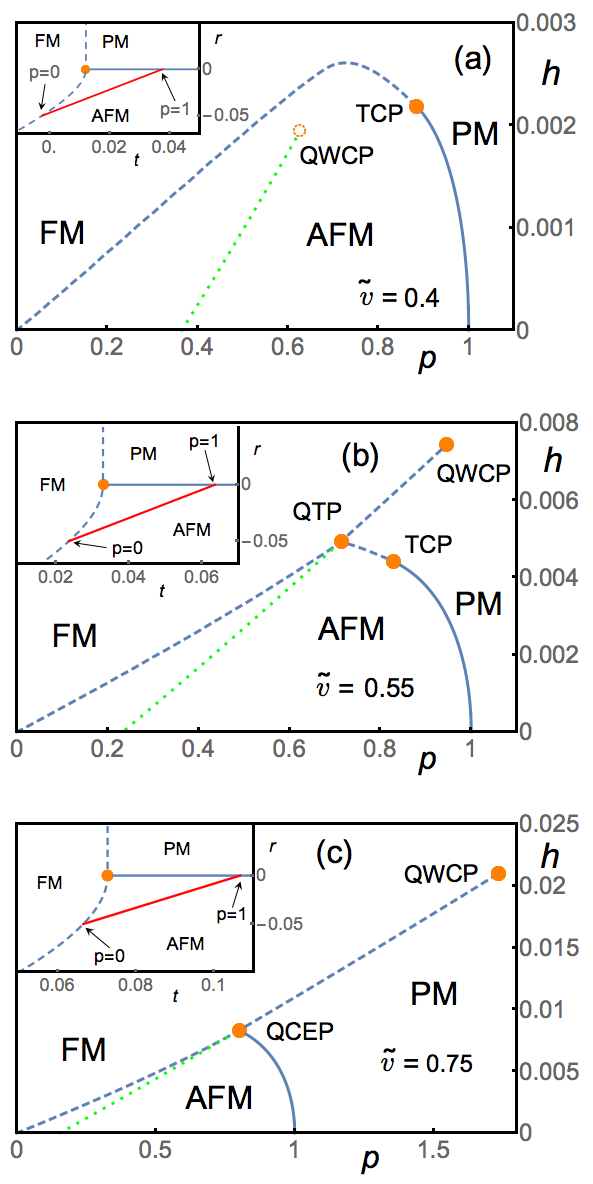}
\caption{Phase diagrams in the $h$-$p$ plane for $u=v=1$, $w=2$, and three different values of the quantum fluctuation parameter ${\tilde v}$.  
              For $\tilde v = 0.5$, panel (a), the structure is qualitatively the same as in the classical MU theory; for larger values
              of $\tilde v$ it is drastically different. 
              Dashed and solid lines denote first and second-order transitions, respectively. The dotted (green) line is the unobservable part of the 
              tricritical wing inside the AFM dome and does not represent a phase transition. Special points are a tricritical point (TCP), 
              a quantum wing critical point (QWCP), a quantum triple point (QTP), and a quantum critical end point (QCEP), see
              see Refs.~\onlinecite{quantum_footnote}, \onlinecite{CEP_footnote} for the nomenclature used. $p$ parameterizes the 
              linear paths in the $r$-$t$ plane, Eq.~(\ref{eq:4}), shown in the insets.}
\label{fig:3}
\end{figure}

{\it $h$-$p$ phase diagrams:} In the presence of a magnetic field $h$, and for relatively large $w$, there are
three possibilities: i) For small $\tilde v$ the QWCP, which marks the end point
of the tricritical wing at $T=0$, lies inside the AFM dome, see Fig.~\ref{fig:3}(a). In this case the wing and
the QWCP are not observable, and the structure of the phase diagram is qualitatively the same as in the classical MU theory:
The AFM dome is delineated on the left by a first-order transition to a (field-polarized) FM state, and on the
right by a second-order transition to a field-polarized PM state, with a tricritical point (TCP) separating the two parts of the dome
boundary. The TCP may lie to the left or to the right of the dome maximum, depending on parameters, see also Fig.~\ref{fig:4}(a)
and the related discussion. ii) For larger values of
$\tilde v$ the QWCP lies outside the AFM dome. If the tricritical wing crosses the dome boundary where the AFM becomes
unstable via a first-order transition, there is a quantum triple point (QTP) where the field-polarized FM and PM phases coexist
with each other and with the AFM phase. The tricritical wing now has a part that is outside of the AFM dome and hence
observable, and the dome boundary consists of three parts: A first-order AFM-FM transition, a first-order AFM-PM transition,
and a second-order AFM-PM transition, with the TCP that also exists in the classical phase diagram separating the latter two.
This case is illustrated in Fig.~\ref{fig:3}(b). iii) For even larger values of $\tilde v$ the tricritical wing intersects the AFM dome
in its second-order section. The dome boundary now consists of only two sections, one first order and one second order, that
are separated by a QCEP, see Fig.~\ref{fig:3}(c). 
In all three cases, the near-linear shape of the left side of the AFM dome reflects the unobservable part of the tricritical wing
inside the dome and thus is a direct consequence of the quantum fluctuations. It is in sharp contrast to the much more symmetric
and evenly curved phase diagram in MU theory. We note in passing that the QWCP and the asymptotic behavior of the
AFM-PM phase boundary near $p=1$ can be determined analytically; the other parts of the phase diagram were obtained by
numerically minimizing the free energy.

For relatively small $w$, there are two possibilities: i) If the path in the $r$-$t$ plane does not cross the FM+AFM phase, 
then the $p$-$h$ phase diagram is qualitatively the same as in the large-$w$ case, see Fig.~\ref{fig:4}(a), which has
the same structure as Fig.~\ref{fig:3}(a). 
ii) If the path does cross the FM+AFM phase, a qualitatively new feature arises: For small external fields, there is a
first-order transition from the FM+AFM phase to the FM phase. This is a true phase transition within the AFM dome 
that has no analog in the large-$w$ case. This line of first-order transitions ends in a quantum critical point (QCP), which in Fig.~\ref{fig:4}(b)
lies within the AFM dome. This is still true for the larger values of $\tilde v$ used in Figs.~\ref{fig:3}(b, c). 
The reason is that with increasing $\tilde v$ the
FM+AFM phase in the $r$-$t$ plane is pushed to larger negative $r$ values. The resulting increase in the $h$-scale 
that characterizes the height of the AFM dome mostly compensates for the increased size of the tricritical wing, and
the first-order transition remains within the dome even for $\tilde v=0.75$. With decreasing $\tilde v$ the
length of the first-order line decreases, and in the classical case it shrinks to zero and the only transition within the dome
is a critical point at $h=0$ \cite{Moriya_Usami_1977}.
\begin{figure}[t]
\includegraphics[width=8cm]{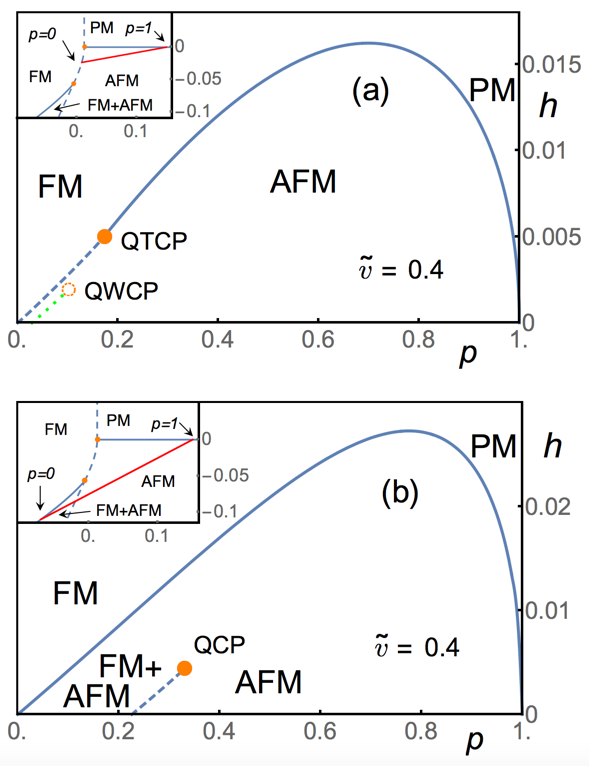}
\caption{Phase diagram in the $h$-$p$ plane for $u=v=1$, $w=0.5$, $\tilde v = 0.4$. The paths parameterized by $p$ are shown 
               in the insets. Solid and dashed lines denote continuous and first-order transition, respectively. QTCP and QCP denote 
               quantum tricritical and quantum critical points, respectively, see see Ref.~\onlinecite{quantum_footnote} for the nomenclature
               used. The dotted (green) line in panel (a) is the unobservable tricritical wing, and QWCP is the unobservable quantum
               wing-critical point. Note the qualitative difference between the two paths.
}
\label{fig:4}
\end{figure}

{\it Relation to Experiments:} Our results help explain why the FM+AFM phase is rarely seen and draws attention when it
is \cite{Pandey_et_al_2013, Ueland_et_al_2015, Fujimori_2015}: It requires a special range of $w$-values
and special properties of the path in the $r$-$t$ plane, since the quantum effects push the FM+AFM phase to 
negative $r$-values of, see Figs.~\ref{fig:1}(b), \ref{fig:4}. 

The observed first-order nature of the FM-AFM transition can be understood even
within the classical MU theory, see Fig.~\ref{fig:1}(a). 
This feature is true {\it a fortiori} in the presence of
quantum fluctuations: The parameter regime where the transition exists is enlarged compared to the classical case, and the transition
is more strongly first order since the FM-PM transition is first order as well. 

A recent experimental study of LaCrGe$_3$ has found an $h$-$p$ phase diagram ($p$ being hydrostatic pressure) consistent with 
Fig.~\ref{fig:3}(c), with a QWCP well outside the AFM dome \cite{Kaluarachchi_et_al_2017}. For CeRuPO a phase diagram
in $T$-$p$-$h$ space has been partially mapped out \cite{Kotegawa_et_al_2013}. In the $h$-$p$ plane a metamagnetic transition 
was found outside the AFM dome that also is consistent with the existence of the
FM-PM first order transition line in Figs.~\ref{fig:3}(b,c). 

{\it Additional remarks:} A nonzero temperature cuts off the nonanalyticities in Eqs.~(\ref{eq:2}, \ref{eq:3}), which leads to new QTCPs
or QCEPs. This 
will be discussed elsewhere \cite{Kirkpatrick_Belitz_2017b}.

In the presence of quenched disorder the nonanalyticities analogous to Eqs.~(\ref{eq:2}, \ref{eq:3})
are stronger and have the opposite sign, and the FM-PM transition is continuous \cite{Brando_et_al_2016a}. It is likely that the FM+AFM
phase is enhanced in this case, but a detailed investigation is necessary.

\medskip
This work was supported by the National Science Foundation under Grant No. DMR-1401449 and Grant No. DMR-1401410. 
Part of this work was performed at the Telluride Science Research Center (TSRC). We thank Valentin Taufour for discussions.

\vskip -0pt

\begin{thebibliography}{23}
\expandafter\ifx\csname natexlab\endcsname\relax\def\natexlab#1{#1}\fi
\expandafter\ifx\csname bibnamefont\endcsname\relax
  \def\bibnamefont#1{#1}\fi
\expandafter\ifx\csname bibfnamefont\endcsname\relax
  \def\bibfnamefont#1{#1}\fi
\expandafter\ifx\csname citenamefont\endcsname\relax
  \def\citenamefont#1{#1}\fi
\expandafter\ifx\csname url\endcsname\relax
  \def\url#1{\texttt{#1}}\fi
\expandafter\ifx\csname urlprefix\endcsname\relax\def\urlprefix{URL }\fi
\providecommand{\bibinfo}[2]{#2}
\providecommand{\eprint}[2][]{\url{#2}}

\bibitem[{\citenamefont{Brando et~al.}(2016)\citenamefont{Brando, Belitz,
  Grosche, and Kirkpatrick}}]{Brando_et_al_2016a}
\bibinfo{author}{\bibfnamefont{M.}~\bibnamefont{Brando}},
  \bibinfo{author}{\bibfnamefont{D.}~\bibnamefont{Belitz}},
  \bibinfo{author}{\bibfnamefont{F.~M.} \bibnamefont{Grosche}},
  \bibnamefont{and} \bibinfo{author}{\bibfnamefont{T.~R.}
  \bibnamefont{Kirkpatrick}}, \bibinfo{journal}{Rev. Mod. Phys.}
  \textbf{\bibinfo{volume}{88}}, \bibinfo{pages}{025006}
  (\bibinfo{year}{2016}).

\bibitem[{\citenamefont{Belitz et~al.}(1999)\citenamefont{Belitz, Kirkpatrick,
  and Vojta}}]{Belitz_Kirkpatrick_Vojta_1999}
\bibinfo{author}{\bibfnamefont{D.}~\bibnamefont{Belitz}},
  \bibinfo{author}{\bibfnamefont{T.~R.} \bibnamefont{Kirkpatrick}},
  \bibnamefont{and} \bibinfo{author}{\bibfnamefont{T.}~\bibnamefont{Vojta}},
  \bibinfo{journal}{Phys. Rev. Lett.} \textbf{\bibinfo{volume}{82}},
  \bibinfo{pages}{4707} (\bibinfo{year}{1999}).

\bibitem[{\citenamefont{Kirkpatrick and
  Belitz}(2012)}]{Kirkpatrick_Belitz_2012b}
\bibinfo{author}{\bibfnamefont{T.~R.} \bibnamefont{Kirkpatrick}}
  \bibnamefont{and} \bibinfo{author}{\bibfnamefont{D.}~\bibnamefont{Belitz}},
  \bibinfo{journal}{Phys. Rev. B} \textbf{\bibinfo{volume}{85}},
  \bibinfo{pages}{134451} (\bibinfo{year}{2012}).

\bibitem[{qua()}]{quantum_footnote}
\bibinfo{note}{Special points in the phase diagram that exist due to quantum
  fluctuations we call ``quantum'' special points and prepend their
  abbreviations with a Q, even if they occur at $T>0$. Special points that
  exist even in a classical Landau theory we denote by their usual names
  without the ``quantum'' designation, regardless of whether or not they exist
  at $T=0$. Note that MU theory, Eq.~(\ref{eq:1}), is perfectly valid, if
  incomplete, at $T=0$.}

\bibitem[{\citenamefont{Belitz et~al.}(2005)\citenamefont{Belitz, Kirkpatrick,
  and Rollb{\"u}hler}}]{Belitz_Kirkpatrick_Rollbuehler_2005}
\bibinfo{author}{\bibfnamefont{D.}~\bibnamefont{Belitz}},
  \bibinfo{author}{\bibfnamefont{T.~R.} \bibnamefont{Kirkpatrick}},
  \bibnamefont{and}
  \bibinfo{author}{\bibfnamefont{J.}~\bibnamefont{Rollb{\"u}hler}},
  \bibinfo{journal}{Phys. Rev. Lett.} \textbf{\bibinfo{volume}{94}},
  \bibinfo{pages}{247205} (\bibinfo{year}{2005}).

\bibitem[{typ()}]{types_of_AFMs_footnote}
\bibinfo{note}{We assume that the FM and AFM orders do not appear
  simultaneously, as they do in ferrimagnets, or in canted AFMs. In these cases
  the (single) QPT is discontinuous by the arguments given in
  Ref.~\onlinecite{Kirkpatrick_Belitz_2012b}.}

\bibitem[{\citenamefont{Muldawer and
  deBergevin}(1961)}]{Muldawer_deBergevin_1961}
\bibinfo{author}{\bibfnamefont{L.}~\bibnamefont{Muldawer}} \bibnamefont{and}
  \bibinfo{author}{\bibfnamefont{F.}~\bibnamefont{deBergevin}},
  \bibinfo{journal}{J. Chem. Phys.} \textbf{\bibinfo{volume}{35}},
  \bibinfo{pages}{1904} (\bibinfo{year}{1961}).

\bibitem[{\citenamefont{Moroni-Klementowicz
  et~al.}(2009)\citenamefont{Moroni-Klementowicz, Brando, Albrecht, Duncan,
  Grosche, Gr\"uner, and Kreiner}}]{Moroni_et_al_2009}
\bibinfo{author}{\bibfnamefont{D.}~\bibnamefont{Moroni-Klementowicz}},
  \bibinfo{author}{\bibfnamefont{M.}~\bibnamefont{Brando}},
  \bibinfo{author}{\bibfnamefont{C.}~\bibnamefont{Albrecht}},
  \bibinfo{author}{\bibfnamefont{W.~J.} \bibnamefont{Duncan}},
  \bibinfo{author}{\bibfnamefont{F.~M.} \bibnamefont{Grosche}},
  \bibinfo{author}{\bibfnamefont{D.}~\bibnamefont{Gr\"uner}}, \bibnamefont{and}
  \bibinfo{author}{\bibfnamefont{G.}~\bibnamefont{Kreiner}},
  \bibinfo{journal}{Phys. Rev. B} \textbf{\bibinfo{volume}{79}},
  \bibinfo{pages}{224410} (\bibinfo{year}{2009}).

\bibitem[{\citenamefont{Kotegawa et~al.}(2013)\citenamefont{Kotegawa, Toyama,
  Kitagawa, Tou, Yamauchi, Matsuoka, and Sugawara}}]{Kotegawa_et_al_2013}
\bibinfo{author}{\bibfnamefont{H.}~\bibnamefont{Kotegawa}},
  \bibinfo{author}{\bibfnamefont{T.}~\bibnamefont{Toyama}},
  \bibinfo{author}{\bibfnamefont{S.}~\bibnamefont{Kitagawa}},
  \bibinfo{author}{\bibfnamefont{H.}~\bibnamefont{Tou}},
  \bibinfo{author}{\bibfnamefont{R.}~\bibnamefont{Yamauchi}},
  \bibinfo{author}{\bibfnamefont{E.}~\bibnamefont{Matsuoka}}, \bibnamefont{and}
  \bibinfo{author}{\bibfnamefont{H.}~\bibnamefont{Sugawara}},
  \bibinfo{journal}{J. Phys. Soc. Jpn.} \textbf{\bibinfo{volume}{82}},
  \bibinfo{pages}{123711} (\bibinfo{year}{2013}).

\bibitem[{\citenamefont{Sidorov et~al.}(2003)\citenamefont{Sidorov, Bauer,
  Frederick, Jeffries, Nakatsuji, Moreno, Thompson, Maple, and
  Fisk}}]{Sidorov_et_al_2003}
\bibinfo{author}{\bibfnamefont{V.~A.} \bibnamefont{Sidorov}},
  \bibinfo{author}{\bibfnamefont{E.~D.} \bibnamefont{Bauer}},
  \bibinfo{author}{\bibfnamefont{N.~A.} \bibnamefont{Frederick}},
  \bibinfo{author}{\bibfnamefont{J.~R.} \bibnamefont{Jeffries}},
  \bibinfo{author}{\bibfnamefont{S.}~\bibnamefont{Nakatsuji}},
  \bibinfo{author}{\bibfnamefont{N.~O.} \bibnamefont{Moreno}},
  \bibinfo{author}{\bibfnamefont{J.~D.} \bibnamefont{Thompson}},
  \bibinfo{author}{\bibfnamefont{M.~B.} \bibnamefont{Maple}}, \bibnamefont{and}
  \bibinfo{author}{\bibfnamefont{Z.}~\bibnamefont{Fisk}},
  \bibinfo{journal}{Phys. Rev. B} \textbf{\bibinfo{volume}{67}},
  \bibinfo{pages}{224419} (\bibinfo{year}{2003}).

\bibitem[{\citenamefont{Enkovaara et~al.}(2003)\citenamefont{Enkovaara, Heczko,
  Ayuela, and Nieminen}}]{Enkovaara_et_al_2003}
\bibinfo{author}{\bibfnamefont{J.}~\bibnamefont{Enkovaara}},
  \bibinfo{author}{\bibfnamefont{O.}~\bibnamefont{Heczko}},
  \bibinfo{author}{\bibfnamefont{A.}~\bibnamefont{Ayuela}}, \bibnamefont{and}
  \bibinfo{author}{\bibfnamefont{R.~M.} \bibnamefont{Nieminen}},
  \bibinfo{journal}{Phys. Rev. Lett.} \textbf{\bibinfo{volume}{67}},
  \bibinfo{pages}{212405} (\bibinfo{year}{2003}).

\bibitem[{\citenamefont{Pandey et~al.}(2013)\citenamefont{Pandey, Ueland,
  Yeninas, Kreyssig, Sapkota, Zhao, Helton, Lynn, McQueeney, Furukawa
  et~al.}}]{Pandey_et_al_2013}
\bibinfo{author}{\bibfnamefont{A.}~\bibnamefont{Pandey}},
  \bibinfo{author}{\bibfnamefont{B.~G.} \bibnamefont{Ueland}},
  \bibinfo{author}{\bibfnamefont{S.}~\bibnamefont{Yeninas}},
  \bibinfo{author}{\bibfnamefont{A.}~\bibnamefont{Kreyssig}},
  \bibinfo{author}{\bibfnamefont{A.}~\bibnamefont{Sapkota}},
  \bibinfo{author}{\bibfnamefont{Y.}~\bibnamefont{Zhao}},
  \bibinfo{author}{\bibfnamefont{J.~S.} \bibnamefont{Helton}},
  \bibinfo{author}{\bibfnamefont{J.~W.} \bibnamefont{Lynn}},
  \bibinfo{author}{\bibfnamefont{R.~J.} \bibnamefont{McQueeney}},
  \bibinfo{author}{\bibfnamefont{Y.}~\bibnamefont{Furukawa}},
  \bibnamefont{et~al.}, \bibinfo{journal}{Phys. Rev. Lett.}
  \textbf{\bibinfo{volume}{111}}, \bibinfo{pages}{047001}
  (\bibinfo{year}{2013}).

\bibitem[{\citenamefont{Kawanaka et~al.}(2010)\citenamefont{Kawanaka, Noguchi,
  Yokoyama, Bando, and Nishihara}}]{Kawanaka_et_al_2009}
\bibinfo{author}{\bibfnamefont{H.}~\bibnamefont{Kawanaka}},
  \bibinfo{author}{\bibfnamefont{A.}~\bibnamefont{Noguchi}},
  \bibinfo{author}{\bibfnamefont{M.}~\bibnamefont{Yokoyama}},
  \bibinfo{author}{\bibfnamefont{H.}~\bibnamefont{Bando}}, \bibnamefont{and}
  \bibinfo{author}{\bibfnamefont{Y.}~\bibnamefont{Nishihara}},
  \bibinfo{journal}{J. Phys. Conf. Series} \textbf{\bibinfo{volume}{200}},
  \bibinfo{pages}{134451} (\bibinfo{year}{2010}).

\bibitem[{\citenamefont{Moriya and Usami}(1977)}]{Moriya_Usami_1977}
\bibinfo{author}{\bibfnamefont{T.}~\bibnamefont{Moriya}} \bibnamefont{and}
  \bibinfo{author}{\bibfnamefont{K.}~\bibnamefont{Usami}},
  \bibinfo{journal}{Solid State Commun.} \textbf{\bibinfo{volume}{23}},
  \bibinfo{pages}{935} (\bibinfo{year}{1977}).

\bibitem[{\citenamefont{Moriya}(1985)}]{Moriya_1985}
\bibinfo{author}{\bibfnamefont{T.}~\bibnamefont{Moriya}},
  \emph{\bibinfo{title}{Spin Fluctuations in Itinerant Electron Magnetism}}
  (\bibinfo{publisher}{Springer, Berlin}, \bibinfo{year}{1985}).

\bibitem[{neg()}]{negative_w_footnote}
\bibinfo{note}{For $u, v >0$ the theory is actually stable for $w >
  -\sqrt{uv}$. Still, $w<0$ makes the omission of terms higher than biquadratic
  in $n$ and $m$ in the free energy questionable.}

\bibitem[{dif()}]{different_AFMs_footnote}
\bibinfo{note}{Experimentally, different AFM phases are sometimes observed with
  increasing $h$ within the AFM dome. These additional phases, if present, do
  not affect the qualitative quantum fluctuation effects discussed here.}

\bibitem[{\citenamefont{Kirkpatrick and
  Belitz}(2017{\natexlab{a}})}]{Kirkpatrick_Belitz_2017b}
\bibinfo{author}{\bibfnamefont{T.~R.} \bibnamefont{Kirkpatrick}}
  \bibnamefont{and} \bibinfo{author}{\bibfnamefont{D.}~\bibnamefont{Belitz}}
  (\bibinfo{year}{2017}{\natexlab{a}}), \bibinfo{note}{unpublished}.

\bibitem[{\citenamefont{Kirkpatrick and
  Belitz}(2017{\natexlab{b}})}]{Kirkpatrick_Belitz_2017}
\bibinfo{author}{\bibfnamefont{T.~R.} \bibnamefont{Kirkpatrick}}
  \bibnamefont{and} \bibinfo{author}{\bibfnamefont{D.}~\bibnamefont{Belitz}},
  \bibinfo{journal}{Phys. Rev. B} \textbf{\bibinfo{volume}{95}},
  \bibinfo{pages}{214401} (\bibinfo{year}{2017}{\natexlab{b}}).

\bibitem[{CEP()}]{CEP_footnote}
\bibinfo{note}{It recently has become common to refer to ordinary critical
  points as ``critical end points''. We use the term in its original meaning,
  i.e., a point where a line of second-order transitions meets two lines of
  first-order transitions.}

\bibitem[{\citenamefont{Ueland et~al.}(2015)\citenamefont{Ueland, Pandey, Lee,
  Sapkota, Choi, Haskel, Rosenberg, Lang, Harmon, Johnston
  et~al.}}]{Ueland_et_al_2015}
\bibinfo{author}{\bibfnamefont{B.~G.} \bibnamefont{Ueland}},
  \bibinfo{author}{\bibfnamefont{A.}~\bibnamefont{Pandey}},
  \bibinfo{author}{\bibfnamefont{Y.}~\bibnamefont{Lee}},
  \bibinfo{author}{\bibfnamefont{A.}~\bibnamefont{Sapkota}},
  \bibinfo{author}{\bibfnamefont{Y.}~\bibnamefont{Choi}},
  \bibinfo{author}{\bibfnamefont{D.}~\bibnamefont{Haskel}},
  \bibinfo{author}{\bibfnamefont{R.~A.} \bibnamefont{Rosenberg}},
  \bibinfo{author}{\bibfnamefont{J.~C.} \bibnamefont{Lang}},
  \bibinfo{author}{\bibfnamefont{B.~N.} \bibnamefont{Harmon}},
  \bibinfo{author}{\bibfnamefont{D.~C.} \bibnamefont{Johnston}},
  \bibnamefont{et~al.}, \bibinfo{journal}{Phys. Rev. Lett.}
  \textbf{\bibinfo{volume}{114}}, \bibinfo{pages}{217001}
  (\bibinfo{year}{2015}).

\bibitem[{Fuj()}]{Fujimori_2015}
\bibinfo{note}{{h}ttps://www.condmatjclub.org/?p=2607}.

\bibitem[{\citenamefont{Kaluarachchi et~al.}()\citenamefont{Kaluarachchi,
  Bud'ko, Canfield, and Taufour}}]{Kaluarachchi_et_al_2017}
\bibinfo{author}{\bibfnamefont{U.~S.} \bibnamefont{Kaluarachchi}},
  \bibinfo{author}{\bibfnamefont{S.~L.} \bibnamefont{Bud'ko}},
  \bibinfo{author}{\bibfnamefont{P.~C.} \bibnamefont{Canfield}},
  \bibnamefont{and} \bibinfo{author}{\bibfnamefont{V.}~\bibnamefont{Taufour}},
  \eprint{{a}rXiv:1611.01212}.

\end{thebibliography}

\end{document}